\documentclass{ws-procs9x6}

\begin{document}

\title{From the Abelian projected flux tube to the dual Abelian Higgs 
model}

\author{E.--M. ILGENFRITZ\footnote{\uppercase{T}his work is done in
collaboration with
\uppercase{Y}.~\uppercase{K}oma,
\uppercase{M}.~\uppercase{K}oma (both
\uppercase{MPI}~\uppercase{M}unich),
and
\uppercase{T}.~\uppercase{S}uzuki
(\uppercase{K}anazawa~\uppercase{U}.)}}

\address{Institut f\"ur Physik, Humboldt Universit\"at zu Berlin, \\
D--12489 Berlin, GERMANY }


\maketitle

\vspace{-6mm}
\abstracts{A recent detailed study of the $Q\bar{Q}$ flux tube
and its DAHM analysis is reported.}
\vspace{-6mm}

In Abelian projected (AP) $SU(2)$ lattice gluodynamics we have
measured the profile of $Q\bar{Q}$ flux tubes for various lattice 
spacings and different $Q\bar{Q}$ distances $R$.
We have fitted the results by the classical flux-tube solution
of the $3D$ lattice $U(1)$ dual Abelian Higgs  model (DAHM)
with external electric charges, taking the finiteness of $R$
into account.\cite{kkis}
On the lattice, the flux-tube profile is characterized by
the correlator of electric field strength and monopole current 
with the (Abelian) Wilson loop.
In the course of work, we have explored and minimized systematic 
effects which could spoil the observed flux-tube profile: a bad 
ground state overlap and the Gribov problem encountered in 
maximal-Abelian gauge fixing.
The superposition of a Coulomb field and a solenoidal field
(as known from DAHM with external sources) has been confirmed on 
the lattice by considering correlators with Wilson loops
expressed in terms of regular and singular parts of the AP gauge 
field, respectively, obtained by Hodge decomposition.
The fits give a dual gauge boson mass $m_{B}$=1091(7) MeV 
and a Higgs mass $m_{\chi}$=953(20) MeV independently of the 
flux-tube length. The Ginzburg-Landau parameter of the gluodynamic
vacuum $\kappa=m_{\chi}/m_{B}= 0.87(2) < 1$  indicates a weak 
type-I superconductor.
In contrast to the masses, the dual gauge coupling $g_m$ is 
obtained $R$-dependent which reflects the antiscreening of the 
non-Abelian quark charge.
Up to $R=0.5$ fm we have found no growing of the flux-tube width, {\it i.e.}, 
no string roughening.

\end{document}